\documentclass[sigconf]{acmart}

\copyrightyear{2026}
\acmYear{2026}
\setcopyright{cc}
\setcctype{by}
\acmConference[RecSys '26]{20th ACM Conference on Recommender Systems}{September 27--October 02, 2026}{Minneapolis, MN, USA}
\acmBooktitle{20th ACM Conference on Recommender Systems (RecSys '26), September 27--October 02, 2026, Minneapolis, MN, USA}
\acmDOI{10.1145/3773078.3831923}
\acmISBN{979-8-4007-2284-4/26/09}

\usepackage{subcaption}
\begin{document}

\title[Mend the Measurement Gap]{Mend the Measurement Gap: Latent User Preference Modeling for Short-Form Video Recommendation}

\author{Shuo Chang}
\orcid{0009-0004-2345-1186}
\affiliation{%
  \institution{Google}
  \city{Mountain View}
  \state{CA}
  \country{USA}
}
\email{changshuo@google.com}

\author{Yueqi Wang}
\orcid{0000-0001-8828-3516}
\affiliation{%
  \institution{Google}
  \city{New York City}
  \state{NY}
  \country{USA}
}
\email{yueqiw@google.com}

\author{Zihuan Diao}
\orcid{0009-0008-9220-783X}
\affiliation{%
  \institution{Google}
  \city{Mountain View}
  \state{CA}
  \country{USA}
}
\email{zdiao@google.com}

\author{Ali Montazer}
\orcid{0000-0002-5467-1331}
\affiliation{%
  \institution{Google}
  \city{Mountain View}
  \state{CA}
  \country{USA}
}
\email{alimontazer@google.com}

\author{Jiangguo Zhang}
\orcid{0000-0001-6504-4360}
\affiliation{%
  \institution{Google}
  \city{Mountain View}
  \state{CA}
  \country{USA}
}
\email{jiangguozhang@google.com}

\author{Joyneel Misra}
\orcid{0000-0001-8306-4832}
\affiliation{%
  \institution{Google}
  \city{Mountain View}
  \state{CA}
  \country{USA}
}
\email{jomi@google.com}

\author{Dapeng Hong}
\orcid{0009-0006-4196-2654}
\affiliation{%
  \institution{Google}
  \city{Mountain View}
  \state{CA}
  \country{USA}
}
\email{dapengh@google.com}

\author{Tomer Margolin}
\orcid{0009-0007-0403-9753}
\affiliation{%
  \institution{Google}
  \city{Mountain View}
  \state{CA}
  \country{USA}
}
\email{tmargolin@google.com}

\author{Sourabh Bansod}
\orcid{0009-0000-2154-0530}
\affiliation{%
  \institution{Google}
  \city{Mountain View}
  \state{CA}
  \country{USA}
}
\email{spbansod@google.com}

\author{Ningren Han}
\orcid{0000-0003-4513-8575}
\affiliation{%
  \department{YouTube}
  \institution{Google}
  \city{Mountain View}
  \state{CA}
  \country{USA}
}
\email{peterhan@google.com}

\renewcommand{\shortauthors}{Chang et al.}

\begin{abstract}
  Recommender systems rely heavily on heterogeneous behavioral feedback to infer user preference. Although abundant, these signals are imperfect measurements: the same observed behavior can arise from different underlying states, such as genuine enjoyment, passive consumption, or inattention. The challenge is especially acute in short-form video, where watch-based signals are strongly affected by measurement confounders such as video duration --- the same watch time can imply different levels of preference for videos of different lengths, while ratio-based metrics can systematically favor short videos. As a result, optimizing raw engagement can amplify measurement artifacts rather than improving user value.
  We propose a Factorized Latent Value Model (FLVM) for measuring user preference from heterogeneous behavioral feedback. The model treats observed behaviors as noisy measurements of a low-dimensional, factorized latent value state and uses structured output heads to model heterogeneous feedback signals. A restricted baseline path captures predictable variation from measurement-confounding features such as video duration, user propensity, and session context, while a routed latent path estimates preference-relevant value advantage. The resulting latent value score can be integrated into an existing recommender system as a ranking feature or ranking score. On YouTube Shorts, a major short-form video platform, this model improves offline metrics and lifts a primary viewer enjoyment metric by 2.67\% in online A/B tests.
\end{abstract}

\ccsdesc[500]{Information systems~Recommender systems}
\ccsdesc[500]{Information systems~Personalization}
\keywords{preference measurement, latent variable models, feedback debiasing, multi-task recommendation, short-form video recommendation}

\maketitle
\section{Introduction}
Recommender systems cannot observe user preference directly; they can only infer it from behavior. That inference is especially unreliable in a low-friction sequential recommendation feed. In this setting, the user consumes a continuous stream of recommendations through swipes, likes, shares, comments, and creator subscriptions. Since explicit feedback is sparse, production systems rely heavily on implicit engagement signals such as watch time and completion ratio~\cite{liu2023immersivefeed,zhao2019recommending}.

But behavioral signals are not direct observations of preference~\cite{lv2025utis,zhao2023uncovering}; they are jointly shaped by user preference, content, video duration, session context, and the system policy. The problem is especially acute on such surfaces, in contrast to click-based discovery pages with active selection~\cite{liu2023immersivefeed,wang2025notallimpressions}. The same observed engagement can reflect different underlying states: a user may watch a video for a long time out of genuine enjoyment or mere passive consumption. A short watch may indicate low preference, quick satisfaction, or active skipping in search of something else.

This ambiguity creates a fundamental challenge for ranking optimization. Raw watch time is strongly affected by video duration, while objectives based on the completion ratio (watch time divided by video duration) can introduce the opposite bias toward shorter videos~\cite{zhan2022deconfounding,zheng2022dvr,tang2023cvrdd,liu2026relativeadvantage}. Explicit actions such as likes, dislikes, and shares are stronger preference indicators but sparse and highly user-dependent. A standard multi-task ranker can learn correlations among these feedback labels~\cite{zhao2019recommending,liu2023immersivefeed,ma2018mmoe}, yet it still directly optimizes observed labels whose meanings are duration-dependent, user-dependent, and semantically entangled.

We call this the \emph{measurement gap}: the distance between the behaviors a recommender observes and the underlying user preference it is trying to measure. Prior work has shown that passive-negative feedback, survey responses, and explicit negative actions provide useful evidence beyond positive engagement alone~\cite{pan2023passivenegative,yu2025unifiedsurvey,raju2025negativefeedback}. We treat these heterogeneous signals jointly as noisy measurements of a shared latent value state\footnote{Here, ``value'' denotes the latent construct that observed engagement signals imperfectly measure---not the reinforcement-learning value function or long-term user value.}, introducing an explicit measurement layer between raw behavioral feedback and ranking optimization.

Building on latent variable recommendation and baseline-relative debiasing, we propose a \textbf{Factorized Latent Value Model} (FLVM) as a structured measurement layer for measuring user preference from heterogeneous behavioral feedback. Because observed behaviors are noisy measurements of an unobserved construct, a latent variable model is a natural fit: FLVM uses a supervised Gaussian variational bottleneck~\cite{alemi2017deepvib}, building on a line of latent variable approaches to recommendation~\cite{liang2018multivae,chang2023latentintent}. Given a user-item impression, the model infers a low-dimensional, factorized latent value from serving-time features and sparsely routes its components to decoders for heterogeneous feedback signals. A restricted baseline path, conditioned only on duration, session context, and user propensity, models the portion of observed engagement predictable from measurement-confounding factors; the latent path then explains residual variation across feedback signals.

This model replaces an existing multi-task ranking model in the pre-scoring stage of our deployment; the resulting latents serve as primary inputs to the scoring formula.

The main contributions are:
\begin{enumerate}
    \item We frame preference learning in recommender systems as a measurement problem, where heterogeneous behavioral signals are noisy and confounded observations of latent user preference.
    \item We introduce FLVM, a structured latent measurement layer that combines a restricted baseline over measurement-confounding features with sparsely routed latent decoders over heterogeneous feedback signals.
    \item We instantiate FLVM in a production short-form video recommender, using duration, session context, and user propensity as baseline features and watch, engagement, valence, and satisfaction-like labels as feedback heads.
    \item We validate FLVM offline and online on YouTube Shorts, a major short-form video platform, showing improved prediction of key feedback signals and a 2.67\% lift in a primary viewer enjoyment metric with neutral guardrails, with consistent gains carrying over to newly uploaded content.
\end{enumerate}

\section{Related Work}

\subsubsection*{Behavior as a noisy measure of preference}
Recent industry work distinguishes observed engagement from the preference or satisfaction a recommender seeks to optimize. In short-form video feeds, peak--end-based user-return modeling shows that impressions should not contribute equally to retention~\cite{wang2025notallimpressions}, while direct survey supervision can reveal interest that watch time, likes, and shares miss~\cite{lv2025utis}. Survey labels are themselves sparse and selected: response-propensity modeling improves satisfaction estimation under survey response bias~\cite{christakopoulou2020deconfounding}, and multi-head survey models capture satisfaction and negative experiences in production~\cite{yu2025unifiedsurvey}. Skips and explicit negative-feedback controls provide complementary evidence beyond positive engagement~\cite{pan2023passivenegative,raju2025negativefeedback}. FLVM shares this measurement motivation but treats these impression-level signals jointly as noisy indicators of a structured latent state, rather than optimizing a single survey or retention target.

\subsubsection*{Watch-time and feedback debiasing}
Although widely used as a supervision signal, watch time is shaped by video duration and does not always reflect user interest. Prior work uses duration-stratified or causal labels~\cite{zhan2022deconfounding,zhang2023leveraging}, separates interest-driven watch time from time spent before users recognize disinterest~\cite{zhao2023uncovering}, and develops architectural and counterfactual methods that suppress duration information or separate its effects~\cite{zheng2022dvr,tang2023cvrdd,zhao2024counteracting}. CVRDD removes duration's direct effect through counterfactual inference, whereas the later CWM formulation treats observed watch time as a duration-truncated realization of latent counterfactual watch time. Most closely related to our baseline-relative construction, Relative Advantage Debiasing maps watch time against user- and item-conditioned reference distributions to create a relative preference target~\cite{liu2026relativeadvantage}. AlignPxtr aligns continuous and discrete behavior distributions across multiple bias dimensions~\cite{lin2025alignpxtr}. Bucketed and prototype-based models further motivate structured watch-time outputs~\cite{sun2024cread,cui2025prowtp}. FLVM instead uses a restricted baseline over duration, session context, and user propensity, and learns a residual variational latent jointly across heterogeneous feedback signals.

\subsubsection*{Variational and multi-task recommendation}
FLVM's stochastic latent layer follows the supervised variational information bottleneck formulation~\cite{alemi2017deepvib}. In recommendation, Multi-VAE adapts variational inference to implicit collaborative filtering, while MacridVAE learns disentangled intent representations~\cite{liang2018multivae,ma2019macridvae}. Multi-Behavior Alignment treats universal user preference as a latent variable inferred from several behavior distributions, and BVAE combines behavior-aware variational representations with multi-task outputs~\cite{xin2023mba,rao2023bvae}. Chang et al.~\cite{chang2023latentintent} provide a production precedent for injecting VAE-inferred user intent into a sequential recommender. Production rankers otherwise commonly predict several feedback objectives directly using shared multi-task architectures~\cite{zhao2019recommending,liu2023immersivefeed,ma2018mmoe}. FLVM differs by learning an impression-level, semantically routed bottleneck from heterogeneous measurements and pairing it with a stopped, restricted baseline before downstream scoring.

\section{Factorized Latent Value Model}
\begin{figure*}
  \centering
  \includegraphics[width=\textwidth]{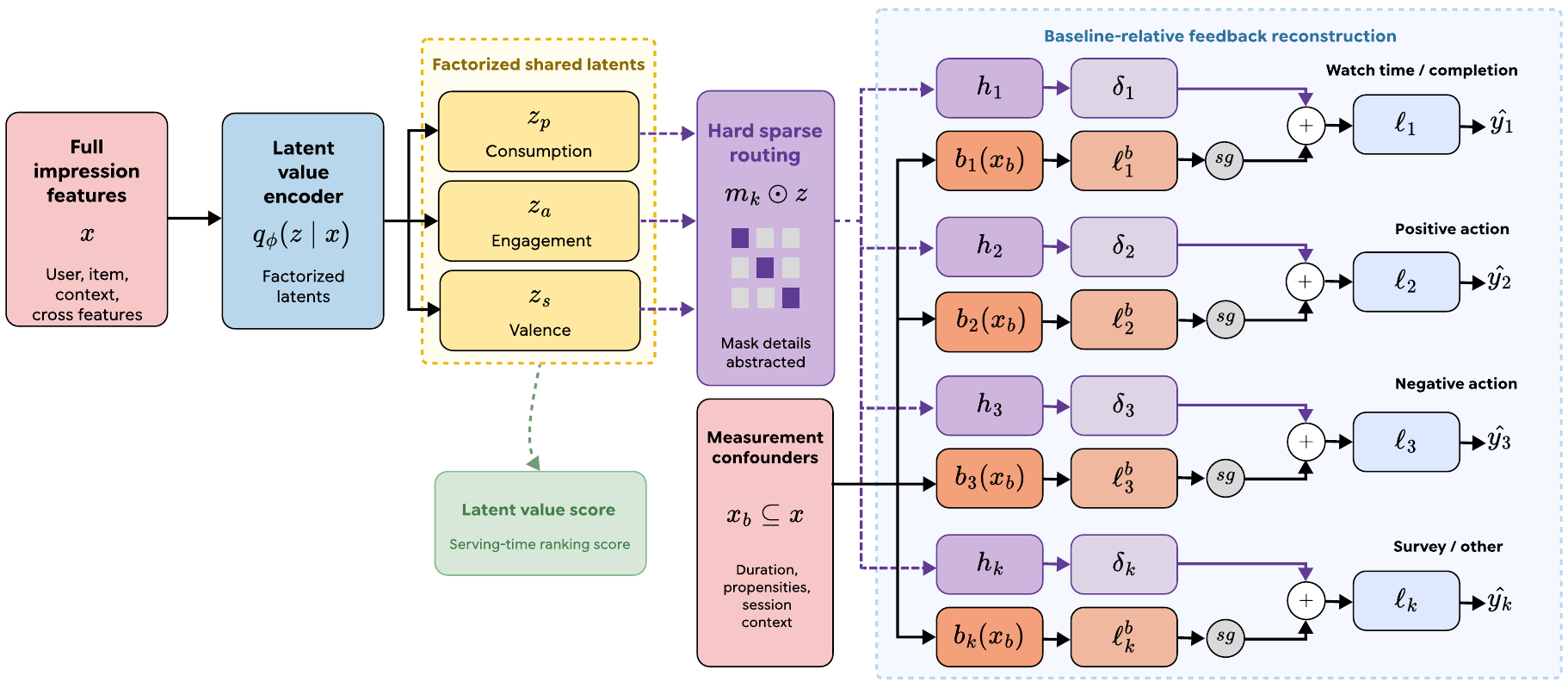}
  \caption{
    Factorized Latent Value Model. The encoder maps full impression features $x$ to a factorized latent value state $z=[z_p,z_a,z_s]$. For each feedback reconstruction task $k$, a task-specific hard sparse mask $m_k$ restricts the latent input to decoder $h_k$, which produces a logit-scale advantage $\delta_k$. In parallel, a restricted baseline $b_k$ uses only measurement-confounding features $x_b\subseteq x$ to produce $\ell_k^b$. The final logit is $\ell_k=\mathrm{sg}(\ell_k^b)+\delta_k$, and predictions are $\hat y_k=o_k(\ell_k)$. The exact mask configuration is abstracted; the latent value score is scalarized from $z$ for serving-time ranking.}
    \Description{
    Architecture diagram of the Factorized Latent Value Model. Full impression features feed a latent value encoder, which produces three latent factors for consumption, engagement, and valence. A hard sparse routing module sends the latent state to several task-specific feedback reconstruction heads, while a restricted set of measurement-confounding features feeds matching baseline heads. Each reconstruction row combines a routed latent advantage with the corresponding stopped baseline logit to produce a final logit and prediction. A separate latent value score is derived from the factorized latent state for serving-time ranking.}
  \label{fig:model}
\end{figure*}

\subsection{Problem Formulation}
The measurement gap motivates treating observed behaviors as noisy measurements rather than direct preference labels. We call the unobserved construct underlying these behaviors \emph{latent value}; because it carries no label of its own, the model must infer it rather than optimize it directly. It is impression-level, capturing the preference-relevant signal for a specific user-item pairing in context.

For each impression, let \(x\) denote the serving-time feature vector and let \(x_b \subseteq x\) denote a subset of measurement-confounding features. In our implementation, \(x_b\) includes video duration, user propensity, and session context. Let
\begin{equation}
    y = \{y_1, y_2, \ldots, y_K\}
\end{equation}
denote the \(K\) heterogeneous feedback signals observed after the impression, such as watch-time buckets,
completion, likes, shares, negative actions, and survey-style feedback.

As shown in Figure~\ref{fig:model}, the model has the dependency structure
\begin{equation}
    x \rightarrow z, \qquad (x_b, z) \rightarrow y
\end{equation}
The encoder defines a conditional distribution \(q_\phi(z \mid x)\) over the latent state. Given \(z\) and \(x_b\), the \(K\) feedback heads are conditionally independent, and each conditional \(p(y_k \mid z, x_b)\) is parameterized by combining two structural paths: a \emph{restricted baseline path} \(b_k\) given only the confounding features \(x_b\), and a \emph{latent path} \(h_k\) given the latent state \(z\). This split is the architectural inductive bias that routes predictable measurement-driven variation to the baseline and lets \(z\) carry preference-relevant residual; the two contributions combine in logit space and pass through a task-specific output transformation \(o_k\):
\begin{equation}
    \ell_k = f_k\big(b_k(x_b),\, h_k(z)\big),
    \qquad
    \hat{y}_k = o_k(\ell_k)
\end{equation}
where \(o_k\) is a sigmoid for binary feedback labels and a softmax for categorical labels. The encoder sees \(x\) in full, so \(z\) can in principle encode any function of \(x\) including \(x_b\); the training scheme encourages the latent path to focus on variation in \(y_k\) that the baseline cannot already explain, so \(z\) carries residual variation in practice. The next sections specialize \(f_k\) and \(h_k\), introduce the structure of \(z\) (factorization and routing), and discuss identifiability, training, and serving.

\subsection{Baseline and Latent Value Advantage}
Our feedback targets are binary or bucketed classification labels, so each decoder operates in logit space. The baseline path is implemented as an MLP; the latent path applies a strictly positive gain to the routed latent state.

The restricted baseline path produces a baseline logit
\begin{equation}
    \ell^b_k = b_k(x_b)
\end{equation}
Because \(x_b\) excludes user-video matching features, the baseline has no direct path to user-item relevance.

The latent path produces a logit-scale \textbf{latent value advantage}
\begin{equation}
    \delta_k = \gamma_k^+ \cdot h_k(z)
\end{equation}
where $h_k$ depends only on the components of $z$ routed to task $k$ (Section~\ref{sec:factors}) and \(\gamma_k^+ \equiv \mathrm{softplus}(\alpha_k) > 0\) is a learned per-signal gain. This positivity stabilizes the sign conventions of $z$ used for identifiability (Section~\ref{sec:identifiability}).

The final decoder logit is
\begin{equation}
    \ell_k = \operatorname{sg}(\ell^b_k) + \delta_k
\end{equation}
where $\operatorname{sg}(\cdot)$ denotes the stop-gradient operator. This baseline-relative formulation is conceptually related to relative-advantage approaches for watch-time debiasing~\cite{liu2026relativeadvantage}, but applies the advantage idea to residuals across heterogeneous feedback signals.

Because gradients are stopped through \(\ell^b_k\), the main feedback objective does not backpropagate into the baseline path: the baseline is updated exclusively from gradients of \(\mathcal{L}_{\mathrm{base}}\) and the main loss rewards \(\delta_k\) only for the portion of \(y_k\) not already explained by the baseline. This encourages confounding effects to be captured by the baseline, yielding a debiased residual.

An alternative is a decoder \(p(y_k \mid z, x_b)\) conditioned directly on the baseline features. Without the stop-gradient and auxiliary baseline loss, the decoder can freely mix \(x_b\) and \(z\), giving \(z\) no reason to encode only residual variation.

\subsection{Factorized Latent Value Variables}
\label{sec:factors}

We factorize the latent value into three components capturing consumption, active engagement, and valence:
\begin{equation}
    z=[z_p, z_a, z_s]
\end{equation}
where \(z_p\) captures consumption-related evidence, \(z_a\) captures active response propensity, and \(z_s\) captures the valence of that engagement: the sentiment-like positive or negative polarity expressed through feedback labels. The valence latent \(z_s\) enters jointly with \(z_a\) to explain feedback such as deep engagement, negative feedback, and satisfaction surveys, where the same active behavior can reflect either a positive or negative reaction.

This factorization reflects the observation that different user feedback signals tell different stories about the same view. A user might watch a long video to completion and scroll on; another might watch and like it; a third might watch and hit the dislike button. On consumption alone, these views are indistinguishable; only the direction of engagement separates the second from the third, and only whether the user actively responded separates the first from the rest. A single unconstrained latent forces the model to average across these qualitatively different signals.

To align each decoder with semantically relevant parts of the latent state, we restrict its latent path with a per-signal hard sparse routing mask $m_k$,
\begin{equation}
    \delta_k = \gamma_k^+ \cdot h_k(m_k \odot z)
\end{equation}
The mask limits each feedback decoder to a small subset of the factorized latent value state, letting related feedback tasks share explanatory factors while preventing unrelated tasks from freely using all latent dimensions. Routing follows the factorization: consumption-related labels are explained through $z_p$, while explicit feedback is explained through $z_a$ and $z_s$ jointly.

This structural constraint also gives the latent components more stable semantics: in an unconstrained bottleneck, latent dimensions can rotate or mix without changing prediction quality, whereas here the meaning of each component is induced by the semantic groups of decoders it is allowed to influence.

\subsection{Inductive Bias for Semantic Roles}
\label{sec:identifiability}

The semantic labels we attach to \(z_p\), \(z_a\), and \(z_s\) are not enforced by the encoder or the standard normal prior alone. A factorized Gaussian latent with an isotropic prior is rotation- and sign-symmetric: nothing in the encoder or the regularization objective prevents the model from arriving at an arbitrary rotation of the three latent axes, in which case the labels ``consumption,'' ``engagement,'' and ``valence'' would be meaningless.

Three architectural choices together reduce this symmetry. First, the routing masks \(m_k\) give each latent influence over a distinct set of decoders. A rotation that mixes \(z_p\) into \(z_s\) would misroute consumption variation into decoders designed to predict valence-signed feedback, increasing the loss. Second, the latent path scales each routed factor by the strictly positive \(\gamma_k^+\), so within the set of decoders sharing a factor, the factor's sign cannot be inverted without loss. Third, the polarity of valence-signed feedback labels, such as explicit satisfaction surveys and negative-engagement signals, anchors the sign convention of \(z_s\). Watch-time label monotonicity and binary engagement intensities play analogous roles for \(z_p\) and \(z_a\).

Together, these three mechanisms supply the inductive bias that encourages stable semantic roles for \(z_p\), \(z_a\), and \(z_s\). Such biases are necessary: Locatello et al.~\cite{locatello2019challenging} show that without structural assumptions on model or data, disentangled representations cannot be recovered unsupervised. Our architecture provides this bias through decoder assignment and sign-anchored labels, rather than through distributional assumptions. Empirically, liked impressions and impressions carrying explicit negative feedback are most separated along the valence factor \(z_s\), with the two groups 1.7--2.0 standard deviations apart across three models independently trained from scratch. This alignment is stable across training: the sign and the ordering of the three factors are identical in every run, with neutral topline metrics in online A/A testing.

\subsection{Training Objective}

We train the model with three components: a main feedback loss, an auxiliary baseline anchoring loss, and a KL regularization loss. We use a supervised Gaussian variational bottleneck with reparameterized sampling and KL regularization~\cite{alemi2017deepvib,kingma2014autoencoding}. The encoder is an MLP with a shared trunk and separate output projections producing a diagonal Gaussian latent value distribution from serving-time features:
\begin{equation}
    q_\phi(z \mid x)=
    \mathcal{N}\left(\mu_\phi(x),\mathrm{diag}(\sigma_\phi^2(x))\right)  
\end{equation}
regularized toward a standard normal prior:
\begin{equation}
    p(z)=\mathcal{N}(0,I)
\end{equation}

\subsubsection*{Main feedback loss}
The main loss trains the encoder and routed latent decoders to predict post-impression feedback:
\begin{equation}
    \mathcal{L}_{\mathrm{main}}=
    \mathbb{E}_{z \sim q_\phi(z \mid x)}
    \left[
    \sum_{k=1}^{K}
    r_k\,\lambda_k\,
    \mathrm{CE}_k
    \left(
    y_k,
    \hat{y}_k
    \right)
    \right]
\end{equation}
where \(r_k \in \{0,1\}\) is an observation mask indicating whether feedback label \(k\) is valid for the impression, and $\lambda_k$ is the task weight balancing the heterogeneous signals. This is important for sparse feedback, which should not be treated as implicit negatives when missing. Since $\ell_k = \operatorname{sg}(\ell^b_k) + \delta_k$, this loss updates the encoder and routed latent decoders, but not the baseline path.

\subsubsection*{Auxiliary baseline anchoring loss}
The baseline path is trained separately with the same observed labels, but only on \(x_b\):
\begin{equation}
    \mathcal{L}_{\mathrm{base}}=
    \sum_{k=1}^{K}r_k\,\lambda_k\,\mathrm{CE}_k\left(y_k,\hat{y}^b_k\right)
\end{equation}
where $\hat{y}^b_k = o_k(\ell^b_k)$ is the baseline prediction and $\lambda_k$ is the per-task weight.

\subsubsection*{Latent regularization}
We regularize the latent value distribution with
\begin{equation}
    \mathcal{L}_{\mathrm{KL}}=
    D_{\mathrm{KL}}\left(q_\phi(z \mid x)\parallel\mathcal{N}(0,I)\right)
\end{equation}
This term keeps the latent space centered and calibrated, reducing the risk that the latent variables become arbitrary task-specific hidden units.

The final training objective is
\begin{equation}
    \mathcal{L}=
    \mathcal{L}_{\mathrm{main}}
    +\mathcal{L}_{\mathrm{base}}
    +\beta \mathcal{L}_{\mathrm{KL}}
\end{equation}
\subsection{Ranking Integration}
At serving time, we scalarize the latent \(z\) into a per-item latent value score using \(\sigma(z_s)\) and \(\mathrm{softplus}(z_p)\). We omit $z_a$ because it captures active response propensity, while $z_s$ carries the direction of that response. Given that propensity without polarity is not evidence of preference, $z_a$ serves only as a training-time auxiliary construct. The resulting score is a primary input to our ranking formula, replacing the existing multi-task pre-scoring stage.

\section{Experiments}
\begin{figure*}[t]
  \centering
  \begin{subfigure}[t]{0.48\textwidth}
    \includegraphics[width=\linewidth]{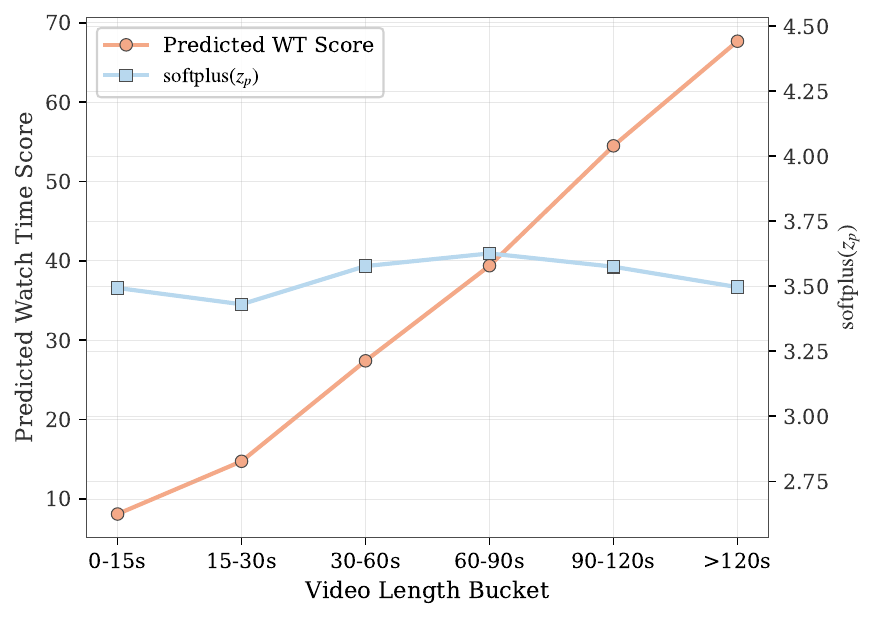}
    \caption{Duration debiasing.}
    \label{fig:duration}
  \end{subfigure}
  \hfill
  \begin{subfigure}[t]{0.48\textwidth}
    \includegraphics[width=\linewidth]{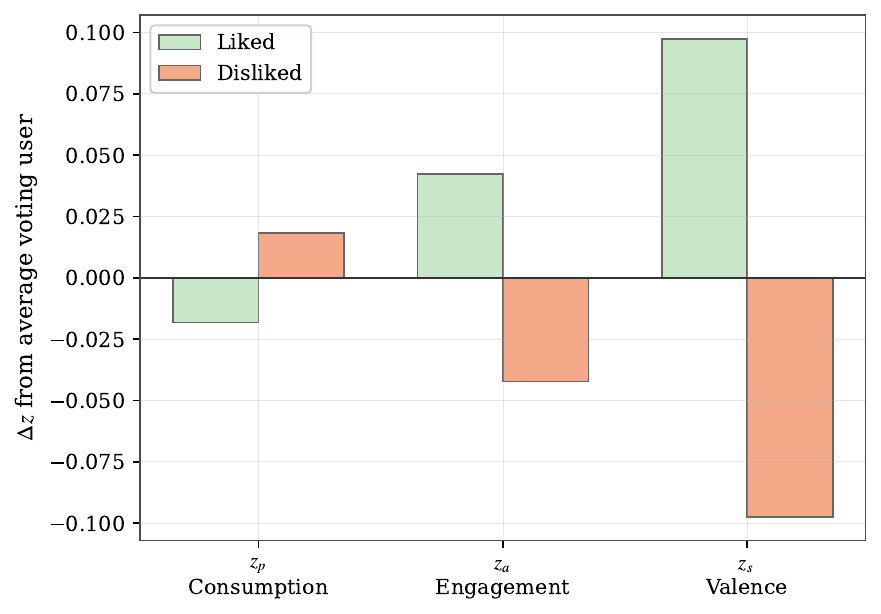}
    \caption{Semantic separation.}
    \label{fig:semantic}
  \end{subfigure}
  \caption{Empirical validation of the two main design choices. (a) The baseline path absorbs duration confounding: predicted item watch time increases with video duration, while the served consumption score $\mathrm{softplus}(z_p)$ stays roughly flat across duration buckets. (b) The three latent factors separate semantically: liked and disliked impressions are most separated along the valence factor $z_s$, less so along the active-engagement factor $z_a$, and approximately invariant on the consumption factor $z_p$.}
  \Description{Two side-by-side charts. (a) Line plot showing predicted item watch time increasing approximately linearly across six video length buckets from 0-15 seconds to over 120 seconds, while softplus(z_p) remains roughly flat at around 3.5 across the same buckets. (b) Bar chart of three latent factors (consumption z_p, engagement z_a, valence z_s), each with two bars showing the relative shift of liked versus disliked impressions from the average user. The valence factor z_s shows the largest separation (positive for liked, negative for disliked); engagement z_a shows moderate separation in the same direction; consumption z_p is approximately invariant.}
    \label{fig:empirical}
\end{figure*}

We evaluate the model on YouTube Shorts, a major short-form video platform with a billion-scale active user base. We compare against the production pre-scoring ranking model, trained with standard multi-task objectives and a post-hoc score combination stage. Figure~\ref{fig:empirical} shows offline validation of the model's two main design choices.

\begin{table}[h]
  \centering
  \caption{Offline predictive performance versus the production multi-task baseline. PR-AUC (higher is better) for binary signals; RMSE (lower is better) for item watch time.}
  \label{tab:offline_metrics}
  \renewcommand{\arraystretch}{1.2}
  \begin{tabular}{l c c c}
    \toprule
    \textbf{Signal} & \textbf{Metric} & \textbf{Production} & \textbf{FLVM} \\
    \midrule
    Item watch time & RMSE $\downarrow$ & 24.47  & \textbf{23.35} \\
    \midrule
    Completion & PR-AUC $\uparrow$ & \textbf{0.6898} & 0.6722 \\
    \midrule
    Like & PR-AUC $\uparrow$ & 0.2522 & \textbf{0.3190} \\
    \midrule
    Dislike & PR-AUC $\uparrow$ & 0.0038 & \textbf{0.0054} \\
    \bottomrule
  \end{tabular}
\end{table}

Table~\ref{tab:offline_metrics} reports offline metrics across core feedback signals. Skip is omitted from the cross-system comparison because the two systems use different skip-label definitions and are not directly comparable. FLVM reduces item watch-time RMSE and improves Like PR-AUC and Dislike PR-AUC, while Completion sees a slight overall drop. However, a fine-grained duration-sliced analysis across all metrics reveals a consistent debiasing effect: FLVM systematically outperforms the baseline on longer videos ($>$60s) across the board—improving Completion PR-AUC by an average of 0.0032—while scoring lower on shorter videos ($<$60s) by an average of 0.0053. Since the production baseline often inflates overall metrics by exploiting short-video heuristics (e.g., inherently higher completion rates), FLVM's performance shift toward harder-to-consume long content indicates a much better capture of genuine user preference. These improvements suggest that jointly explaining heterogeneous feedback through a compact latent value state denoises sparse and confounded signals more effectively than per-task heads.

\begin{table}[t]
  \centering
  \caption{Decomposition: PR-AUC of each path in FLVM. Baseline-only $= o_k(\ell_k^b)$; Latent-only $= o_k(\delta_k)$; Full $= o_k(\ell_k)$. The two paths together exceed either alone.}
  \label{tab:decomposition}
  \renewcommand{\arraystretch}{1.2}
  \begin{tabular}{l c c c}
    \toprule
    \textbf{Signal} & \textbf{Baseline} & \textbf{Latent} & \textbf{Full} \\
    \midrule
    Completion & 0.3065 & 0.6617 & \textbf{0.6722} \\
    \midrule
    Like & 0.1860 & 0.0715 & \textbf{0.3190} \\
    \midrule
    Skip & 0.1840 & 0.5016 & \textbf{0.5100} \\
    \midrule
    Dislike & 0.0035 & 0.0035 & \textbf{0.0054} \\
    \bottomrule
  \end{tabular}
\end{table}

Table~\ref{tab:decomposition} decomposes FLVM into its two paths. For all signals, the Full model exceeds either path alone, confirming that the baseline and latent contributions are complementary. However, the exact dependency varies by signal and item lifecycle. Notably, we observe a powerful synergistic effect on active engagement (Like): each path predicts Like poorly on its own, yet together they far outperform either alone. This suggests that explicit actions such as Like are heavily masked by measurement-confounding features (e.g., user propensity), and that accounting for these confounders in the restricted baseline lets the latent path recover the preference-relevant signal. For passive behaviors like Completion and Skip, the latent path drives the majority of the predictive performance, though the baseline remains strictly necessary for full accuracy.

In our online A/B experiment, spanning 14 days at our standard experiment traffic allocation, the latent value model improves over our production baseline multi-task pre-scorer, increasing our primary viewer enjoyment metric by \textbf{2.67\%} with neutral guardrail metrics. To verify the robustness and generalizability of the learned latent value state across different item lifecycles, we evaluated performance on the cold-start slice (fresh videos with limited exposure history). On this group, the model achieved a consistent $>$2\% lift in helping new videos successfully attract early engagement and transition into mainstream distribution. This demonstrates that by separating genuine preference from structural and statistical noise, FLVM avoids overfitting to historic engagement and robustly generalizes to trailing-edge data distributions.

\section{Conclusion}
We presented a Factorized Latent Value Model that treats observed user feedback as noisy measurements of a low-dimensional, factorized latent value state. The model combines three design choices:
\begin{enumerate}
    \item A baseline restricted to measurement-confounding features.
    \item Factorized latent value variables that represent different aspects of user preference.
    \item Predefined hard sparse decoder routing that connects each latent factor only to semantically relevant feedback signals.
\end{enumerate}
The resulting latent value score integrates into an industrial recommender system with meaningful offline and online gains, demonstrating robust generalizability across diverse content lifecycles and varying levels of data sparsity. Empirically, the learned latent factors take on stable semantic roles induced by model structure and feedback signals.

\let\balance\relax
\bibliographystyle{ACM-Reference-Format}
\bibliography{sample-base}


\begin{thebibliography}{27}


\ifx \showCODEN    \undefined \def \showCODEN     #1{\unskip}     \fi
\ifx \showISBNx    \undefined \def \showISBNx     #1{\unskip}     \fi
\ifx \showISBNxiii \undefined \def \showISBNxiii  #1{\unskip}     \fi
\ifx \showISSN     \undefined \def \showISSN      #1{\unskip}     \fi
\ifx \showLCCN     \undefined \def \showLCCN      #1{\unskip}     \fi
\ifx \shownote     \undefined \def \shownote      #1{#1}          \fi
\ifx \showarticletitle \undefined \def \showarticletitle #1{#1}   \fi
\ifx \showURL      \undefined \def \showURL       {\relax}        \fi
\providecommand\bibfield[2]{#2}
\providecommand\bibinfo[2]{#2}
\providecommand\natexlab[1]{#1}
\providecommand\showeprint[2][]{arXiv:#2}

\bibitem[Alemi et~al\mbox{.}(2017)]%
        {alemi2017deepvib}
\bibfield{author}{\bibinfo{person}{Alexander~A. Alemi}, \bibinfo{person}{Ian
  Fischer}, \bibinfo{person}{Joshua~V. Dillon}, {and} \bibinfo{person}{Kevin
  Murphy}.} \bibinfo{year}{2017}\natexlab{}.
\newblock \showarticletitle{Deep Variational Information Bottleneck}. In
  \bibinfo{booktitle}{\emph{5th International Conference on Learning
  Representations, ICLR 2017, Conference Track Proceedings}}.
  \bibinfo{publisher}{OpenReview.net}, \bibinfo{address}{Toulon, France},
  \bibinfo{numpages}{14}~pages.
\newblock
\showeprint[arxiv]{1612.00410}


\bibitem[Chang et~al\mbox{.}(2023)]%
        {chang2023latentintent}
\bibfield{author}{\bibinfo{person}{Bo Chang}, \bibinfo{person}{Alexandros
  Karatzoglou}, \bibinfo{person}{Yuyan Wang}, \bibinfo{person}{Can Xu},
  \bibinfo{person}{Ed~H. Chi}, {and} \bibinfo{person}{Minmin Chen}.}
  \bibinfo{year}{2023}\natexlab{}.
\newblock \showarticletitle{Latent User Intent Modeling for Sequential
  Recommenders}. In \bibinfo{booktitle}{\emph{Companion Proceedings of the ACM
  Web Conference 2023}} \emph{(\bibinfo{series}{WWW '23 Companion})}.
  \bibinfo{publisher}{Association for Computing Machinery},
  \bibinfo{address}{New York, NY, USA}, \bibinfo{pages}{427--431}.
\newblock
\showeprint[arxiv]{2211.09832}
\href{https://doi.org/10.1145/3543873.3584641}{doi:\nolinkurl{10.1145/3543873.3584641}}


\bibitem[Christakopoulou et~al\mbox{.}(2020)]%
        {christakopoulou2020deconfounding}
\bibfield{author}{\bibinfo{person}{Konstantina Christakopoulou},
  \bibinfo{person}{Madeleine Traverse}, \bibinfo{person}{Trevor Potter},
  \bibinfo{person}{Emma Marriott}, \bibinfo{person}{Daniel Li},
  \bibinfo{person}{Chris Haulk}, \bibinfo{person}{Ed~H. Chi}, {and}
  \bibinfo{person}{Minmin Chen}.} \bibinfo{year}{2020}\natexlab{}.
\newblock \showarticletitle{Deconfounding User Satisfaction Estimation from
  Response Rate Bias}. In \bibinfo{booktitle}{\emph{Proceedings of the 14th ACM
  Conference on Recommender Systems}} \emph{(\bibinfo{series}{RecSys '20})}.
  \bibinfo{publisher}{Association for Computing Machinery},
  \bibinfo{address}{New York, NY, USA}, \bibinfo{pages}{450--455}.
\newblock
\href{https://doi.org/10.1145/3383313.3412208}{doi:\nolinkurl{10.1145/3383313.3412208}}


\bibitem[Cui et~al\mbox{.}(2025)]%
        {cui2025prowtp}
\bibfield{author}{\bibinfo{person}{Chao Cui}, \bibinfo{person}{Shisong Tang},
  \bibinfo{person}{Fan Li}, \bibinfo{person}{Jiechao Gao}, {and}
  \bibinfo{person}{Hechang Chen}.} \bibinfo{year}{2025}\natexlab{}.
\newblock \showarticletitle{Calibrating Video Watch-time Predictions with
  Credible Prototype Alignment}. In \bibinfo{booktitle}{\emph{Proceedings of
  the 42nd International Conference on Machine Learning}},
  Vol.~\bibinfo{volume}{267}. \bibinfo{publisher}{PMLR},
  \bibinfo{address}{Vancouver, Canada}, \bibinfo{pages}{11563--11584}.
\newblock


\bibitem[Kingma and Welling(2014)]%
        {kingma2014autoencoding}
\bibfield{author}{\bibinfo{person}{Diederik~P. Kingma} {and}
  \bibinfo{person}{Max Welling}.} \bibinfo{year}{2014}\natexlab{}.
\newblock \showarticletitle{Auto-Encoding Variational Bayes}. In
  \bibinfo{booktitle}{\emph{2nd International Conference on Learning
  Representations, ICLR 2014, Conference Track Proceedings}}.
  \bibinfo{publisher}{OpenReview.net}, \bibinfo{address}{Banff, AB, Canada},
  \bibinfo{numpages}{14}~pages.
\newblock
\showeprint[arxiv]{1312.6114}


\bibitem[Liang et~al\mbox{.}(2018)]%
        {liang2018multivae}
\bibfield{author}{\bibinfo{person}{Dawen Liang}, \bibinfo{person}{Rahul~G.
  Krishnan}, \bibinfo{person}{Matthew~D. Hoffman}, {and} \bibinfo{person}{Tony
  Jebara}.} \bibinfo{year}{2018}\natexlab{}.
\newblock \showarticletitle{Variational Autoencoders for Collaborative
  Filtering}. In \bibinfo{booktitle}{\emph{Proceedings of The Web Conference
  2018}} \emph{(\bibinfo{series}{WWW '18})}. \bibinfo{publisher}{International
  World Wide Web Conferences Steering Committee}, \bibinfo{address}{Republic
  and Canton of Geneva, CHE}, \bibinfo{pages}{689--698}.
\newblock
\showeprint[arxiv]{1802.05814}
\href{https://doi.org/10.1145/3178876.3186150}{doi:\nolinkurl{10.1145/3178876.3186150}}


\bibitem[Lin et~al\mbox{.}(2025)]%
        {lin2025alignpxtr}
\bibfield{author}{\bibinfo{person}{Chengzhi Lin}, \bibinfo{person}{Chuyuan
  Wang}, \bibinfo{person}{Annan Xie}, \bibinfo{person}{Wuhong Wang},
  \bibinfo{person}{Ziye Zhang}, \bibinfo{person}{Canguang Ruan},
  \bibinfo{person}{Yuancai Huang}, {and} \bibinfo{person}{Yongqi Liu}.}
  \bibinfo{year}{2025}\natexlab{}.
\newblock \bibinfo{title}{{AlignPxtr}: Aligning Predicted Behavior
  Distributions for Bias-Free Video Recommendations}.
\newblock
\showeprint[arxiv]{2503.06920}~[cs.IR]


\bibitem[Liu et~al\mbox{.}(2026)]%
        {liu2026relativeadvantage}
\bibfield{author}{\bibinfo{person}{Emily Liu}, \bibinfo{person}{Kuan Han},
  \bibinfo{person}{Minfeng Zhan}, \bibinfo{person}{Bocheng Zhao},
  \bibinfo{person}{Guanyu Mu}, {and} \bibinfo{person}{Yang Song}.}
  \bibinfo{year}{2026}\natexlab{}.
\newblock \showarticletitle{Relative Advantage Debiasing for Watch-Time
  Prediction in Short-Video Recommendation}.
\newblock \bibinfo{journal}{\emph{Proceedings of the AAAI Conference on
  Artificial Intelligence}} \bibinfo{volume}{40}, \bibinfo{number}{18}
  (\bibinfo{year}{2026}), \bibinfo{pages}{15296--15305}.
\newblock
\showeprint[arxiv]{2508.11086}
\href{https://doi.org/10.1609/aaai.v40i18.38555}{doi:\nolinkurl{10.1609/aaai.v40i18.38555}}


\bibitem[Liu et~al\mbox{.}(2023)]%
        {liu2023immersivefeed}
\bibfield{author}{\bibinfo{person}{Qingyun Liu}, \bibinfo{person}{Zhe Zhao},
  \bibinfo{person}{Liang Liu}, \bibinfo{person}{Zhen Zhang},
  \bibinfo{person}{Junjie Shan}, \bibinfo{person}{Yuening Li},
  \bibinfo{person}{Shuchao Bi}, \bibinfo{person}{Lichan Hong}, {and}
  \bibinfo{person}{Ed~H. Chi}.} \bibinfo{year}{2023}\natexlab{}.
\newblock \showarticletitle{Multitask Ranking System for Immersive Feed and No
  More Clicks: A Case Study of Short-Form Video Recommendation}. In
  \bibinfo{booktitle}{\emph{Proceedings of the 32nd ACM International
  Conference on Information and Knowledge Management}}
  \emph{(\bibinfo{series}{CIKM '23})}. \bibinfo{publisher}{Association for
  Computing Machinery}, \bibinfo{address}{New York, NY, USA},
  \bibinfo{pages}{4709--4716}.
\newblock
\href{https://doi.org/10.1145/3583780.3615489}{doi:\nolinkurl{10.1145/3583780.3615489}}


\bibitem[Locatello et~al\mbox{.}(2019)]%
        {locatello2019challenging}
\bibfield{author}{\bibinfo{person}{Francesco Locatello},
  \bibinfo{person}{Stefan Bauer}, \bibinfo{person}{Mario Lucic},
  \bibinfo{person}{Gunnar R{\"a}tsch}, \bibinfo{person}{Sylvain Gelly},
  \bibinfo{person}{Bernhard Sch{\"o}lkopf}, {and} \bibinfo{person}{Olivier
  Bachem}.} \bibinfo{year}{2019}\natexlab{}.
\newblock \showarticletitle{Challenging Common Assumptions in the Unsupervised
  Learning of Disentangled Representations}. In
  \bibinfo{booktitle}{\emph{Proceedings of the 36th International Conference on
  Machine Learning}} \emph{(\bibinfo{series}{ICML '19},
  Vol.~\bibinfo{volume}{97})}. \bibinfo{publisher}{PMLR},
  \bibinfo{address}{Long Beach, California, USA}, \bibinfo{pages}{4114--4124}.
\newblock
\showeprint[arxiv]{1811.12359}


\bibitem[Lv et~al\mbox{.}(2025)]%
        {lv2025utis}
\bibfield{author}{\bibinfo{person}{Mengxi Lv}, \bibinfo{person}{Drew Hogg},
  \bibinfo{person}{Thomas Grubb}, \bibinfo{person}{Shashank Bassi},
  \bibinfo{person}{Min Li}, \bibinfo{person}{Cayman Simpson}, {and}
  \bibinfo{person}{Senthil Rajagopalan}.} \bibinfo{year}{2025}\natexlab{}.
\newblock \showarticletitle{Improve the Personalization of Large-Scale Ranking
  Systems by Integrating User Survey Feedback}. In
  \bibinfo{booktitle}{\emph{Proceedings of the Nineteenth ACM Conference on
  Recommender Systems}} \emph{(\bibinfo{series}{RecSys '25})}.
  \bibinfo{publisher}{Association for Computing Machinery},
  \bibinfo{address}{New York, NY, USA}, \bibinfo{pages}{971--974}.
\newblock
\href{https://doi.org/10.1145/3705328.3748119}{doi:\nolinkurl{10.1145/3705328.3748119}}


\bibitem[Ma et~al\mbox{.}(2018)]%
        {ma2018mmoe}
\bibfield{author}{\bibinfo{person}{Jiaqi Ma}, \bibinfo{person}{Zhe Zhao},
  \bibinfo{person}{Xinyang Yi}, \bibinfo{person}{Jilin Chen},
  \bibinfo{person}{Lichan Hong}, {and} \bibinfo{person}{Ed~H. Chi}.}
  \bibinfo{year}{2018}\natexlab{}.
\newblock \showarticletitle{Modeling Task Relationships in Multi-task Learning
  with Multi-gate Mixture-of-Experts}. In \bibinfo{booktitle}{\emph{Proceedings
  of the 24th ACM SIGKDD International Conference on Knowledge Discovery \&
  Data Mining}} \emph{(\bibinfo{series}{KDD '18})}.
  \bibinfo{publisher}{Association for Computing Machinery},
  \bibinfo{address}{New York, NY, USA}, \bibinfo{pages}{1930--1939}.
\newblock
\href{https://doi.org/10.1145/3219819.3220007}{doi:\nolinkurl{10.1145/3219819.3220007}}


\bibitem[Ma et~al\mbox{.}(2019)]%
        {ma2019macridvae}
\bibfield{author}{\bibinfo{person}{Jianxin Ma}, \bibinfo{person}{Chang Zhou},
  \bibinfo{person}{Peng Cui}, \bibinfo{person}{Hongxia Yang}, {and}
  \bibinfo{person}{Wenwu Zhu}.} \bibinfo{year}{2019}\natexlab{}.
\newblock \showarticletitle{Learning Disentangled Representations for
  Recommendation}. In \bibinfo{booktitle}{\emph{Advances in Neural Information
  Processing Systems 32}}. \bibinfo{publisher}{Curran Associates, Inc.},
  \bibinfo{address}{Red Hook, NY, USA}, \bibinfo{pages}{5712--5723}.
\newblock
\showeprint[arxiv]{1910.14238}


\bibitem[Pan et~al\mbox{.}(2023)]%
        {pan2023passivenegative}
\bibfield{author}{\bibinfo{person}{Yunzhu Pan}, \bibinfo{person}{Chen Gao},
  \bibinfo{person}{Jianxin Chang}, \bibinfo{person}{Yanan Niu},
  \bibinfo{person}{Yang Song}, \bibinfo{person}{Kun Gai},
  \bibinfo{person}{Depeng Jin}, {and} \bibinfo{person}{Yong Li}.}
  \bibinfo{year}{2023}\natexlab{}.
\newblock \showarticletitle{Understanding and Modeling Passive-Negative
  Feedback for Short-Video Sequential Recommendation}. In
  \bibinfo{booktitle}{\emph{Proceedings of the 17th ACM Conference on
  Recommender Systems}} \emph{(\bibinfo{series}{RecSys '23})}.
  \bibinfo{publisher}{Association for Computing Machinery},
  \bibinfo{address}{New York, NY, USA}, \bibinfo{pages}{540--550}.
\newblock
\href{https://doi.org/10.1145/3604915.3608814}{doi:\nolinkurl{10.1145/3604915.3608814}}


\bibitem[Raju et~al\mbox{.}(2025)]%
        {raju2025negativefeedback}
\bibfield{author}{\bibinfo{person}{Madhura Raju}, \bibinfo{person}{Manisha
  Sharma}, \bibinfo{person}{Hongyu Xiong}, \bibinfo{person}{Bingfeng Deng},
  {and} \bibinfo{person}{Meng Na}.} \bibinfo{year}{2025}\natexlab{}.
\newblock \showarticletitle{Leveraging Explicit Negative Feedback in
  Large-Scale Recommendation Systems: A Case Study}. In
  \bibinfo{booktitle}{\emph{Proceedings of the Nineteenth ACM Conference on
  Recommender Systems}} \emph{(\bibinfo{series}{RecSys '25})}.
  \bibinfo{publisher}{Association for Computing Machinery},
  \bibinfo{address}{New York, NY, USA}, \bibinfo{pages}{999--1001}.
\newblock
\href{https://doi.org/10.1145/3705328.3748145}{doi:\nolinkurl{10.1145/3705328.3748145}}


\bibitem[Rao et~al\mbox{.}(2023)]%
        {rao2023bvae}
\bibfield{author}{\bibinfo{person}{Qianzhen Rao}, \bibinfo{person}{Yang Liu},
  \bibinfo{person}{Weike Pan}, {and} \bibinfo{person}{Zhong Ming}.}
  \bibinfo{year}{2023}\natexlab{}.
\newblock \showarticletitle{{BVAE}: Behavior-aware Variational Autoencoder for
  Multi-behavior Multi-task Recommendation}. In
  \bibinfo{booktitle}{\emph{Proceedings of the 17th ACM Conference on
  Recommender Systems}} \emph{(\bibinfo{series}{RecSys '23})}.
  \bibinfo{publisher}{Association for Computing Machinery},
  \bibinfo{address}{New York, NY, USA}, \bibinfo{pages}{625--636}.
\newblock
\href{https://doi.org/10.1145/3604915.3608781}{doi:\nolinkurl{10.1145/3604915.3608781}}


\bibitem[Sun et~al\mbox{.}(2024)]%
        {sun2024cread}
\bibfield{author}{\bibinfo{person}{Jie Sun}, \bibinfo{person}{Zhaoying Ding},
  \bibinfo{person}{Xiaoshuang Chen}, \bibinfo{person}{Qi Chen},
  \bibinfo{person}{Yincheng Wang}, \bibinfo{person}{Kaiqiao Zhan}, {and}
  \bibinfo{person}{Ben Wang}.} \bibinfo{year}{2024}\natexlab{}.
\newblock \showarticletitle{{CREAD}: A Classification-Restoration Framework
  with Error Adaptive Discretization for Watch Time Prediction in Video
  Recommender Systems}.
\newblock \bibinfo{journal}{\emph{Proceedings of the AAAI Conference on
  Artificial Intelligence}} \bibinfo{volume}{38}, \bibinfo{number}{8}
  (\bibinfo{year}{2024}), \bibinfo{pages}{9027--9034}.
\newblock
\showeprint[arxiv]{2401.07521}
\href{https://doi.org/10.1609/aaai.v38i8.28752}{doi:\nolinkurl{10.1609/aaai.v38i8.28752}}


\bibitem[Tang et~al\mbox{.}(2023)]%
        {tang2023cvrdd}
\bibfield{author}{\bibinfo{person}{Shisong Tang}, \bibinfo{person}{Qing Li},
  \bibinfo{person}{Dingmin Wang}, \bibinfo{person}{Ci Gao},
  \bibinfo{person}{Wentao Xiao}, \bibinfo{person}{Dan Zhao},
  \bibinfo{person}{Yong Jiang}, \bibinfo{person}{Qian Ma}, {and}
  \bibinfo{person}{Aoyang Zhang}.} \bibinfo{year}{2023}\natexlab{}.
\newblock \showarticletitle{Counterfactual Video Recommendation for Duration
  Debiasing}. In \bibinfo{booktitle}{\emph{Proceedings of the 29th ACM SIGKDD
  Conference on Knowledge Discovery and Data Mining}}
  \emph{(\bibinfo{series}{KDD '23})}. \bibinfo{publisher}{Association for
  Computing Machinery}, \bibinfo{address}{New York, NY, USA},
  \bibinfo{pages}{4894--4903}.
\newblock
\href{https://doi.org/10.1145/3580305.3599797}{doi:\nolinkurl{10.1145/3580305.3599797}}


\bibitem[Wang et~al\mbox{.}(2025)]%
        {wang2025notallimpressions}
\bibfield{author}{\bibinfo{person}{Yuyan Wang}, \bibinfo{person}{Jing Zhong},
  \bibinfo{person}{Yuxin Cui}, \bibinfo{person}{Zhaohui Guo},
  \bibinfo{person}{Chuanqi Wei}, \bibinfo{person}{Yanchen Wang}, {and}
  \bibinfo{person}{Zellux Wang}.} \bibinfo{year}{2025}\natexlab{}.
\newblock \showarticletitle{Not All Impressions Are Created Equal:
  Psychology-Informed Retention Optimization for Short-Form Video
  Recommendation}. In \bibinfo{booktitle}{\emph{Proceedings of the Nineteenth
  ACM Conference on Recommender Systems}} \emph{(\bibinfo{series}{RecSys
  '25})}. \bibinfo{publisher}{Association for Computing Machinery},
  \bibinfo{address}{New York, NY, USA}, \bibinfo{pages}{1022--1025}.
\newblock
\href{https://doi.org/10.1145/3705328.3748122}{doi:\nolinkurl{10.1145/3705328.3748122}}


\bibitem[Xin et~al\mbox{.}(2023)]%
        {xin2023mba}
\bibfield{author}{\bibinfo{person}{Xin Xin}, \bibinfo{person}{Xiangyuan Liu},
  \bibinfo{person}{Hanbing Wang}, \bibinfo{person}{Pengjie Ren},
  \bibinfo{person}{Zhumin Chen}, \bibinfo{person}{Jiahuan Lei},
  \bibinfo{person}{Xinlei Shi}, \bibinfo{person}{Hengliang Luo},
  \bibinfo{person}{Joemon~M. Jose}, \bibinfo{person}{Maarten de Rijke}, {and}
  \bibinfo{person}{Zhaochun Ren}.} \bibinfo{year}{2023}\natexlab{}.
\newblock \showarticletitle{Improving Implicit Feedback-Based Recommendation
  through Multi-Behavior Alignment}. In \bibinfo{booktitle}{\emph{Proceedings
  of the 46th International ACM SIGIR Conference on Research and Development in
  Information Retrieval}} \emph{(\bibinfo{series}{SIGIR '23})}.
  \bibinfo{publisher}{Association for Computing Machinery},
  \bibinfo{address}{New York, NY, USA}, \bibinfo{pages}{932--941}.
\newblock
\showeprint[arxiv]{2305.05585}
\href{https://doi.org/10.1145/3539618.3591697}{doi:\nolinkurl{10.1145/3539618.3591697}}


\bibitem[Yu et~al\mbox{.}(2025)]%
        {yu2025unifiedsurvey}
\bibfield{author}{\bibinfo{person}{Chenghui Yu}, \bibinfo{person}{Haoze Wu},
  \bibinfo{person}{Jian Ding}, \bibinfo{person}{Bingfeng Deng}, {and}
  \bibinfo{person}{Hongyu Xiong}.} \bibinfo{year}{2025}\natexlab{}.
\newblock \showarticletitle{Unified Survey Modeling to Limit Negative User
  Experiences in Recommendation Systems}. In
  \bibinfo{booktitle}{\emph{Proceedings of the Nineteenth ACM Conference on
  Recommender Systems}} \emph{(\bibinfo{series}{RecSys '25})}.
  \bibinfo{publisher}{Association for Computing Machinery},
  \bibinfo{address}{New York, NY, USA}, \bibinfo{pages}{1104--1107}.
\newblock
\href{https://doi.org/10.1145/3705328.3748108}{doi:\nolinkurl{10.1145/3705328.3748108}}


\bibitem[Zhan et~al\mbox{.}(2022)]%
        {zhan2022deconfounding}
\bibfield{author}{\bibinfo{person}{Ruohan Zhan}, \bibinfo{person}{Changhua
  Pei}, \bibinfo{person}{Qiang Su}, \bibinfo{person}{Jianfeng Wen},
  \bibinfo{person}{Xueliang Wang}, \bibinfo{person}{Guanyu Mu},
  \bibinfo{person}{Dong Zheng}, \bibinfo{person}{Peng Jiang}, {and}
  \bibinfo{person}{Kun Gai}.} \bibinfo{year}{2022}\natexlab{}.
\newblock \showarticletitle{Deconfounding Duration Bias in Watch-Time
  Prediction for Video Recommendation}. In
  \bibinfo{booktitle}{\emph{Proceedings of the 28th ACM SIGKDD Conference on
  Knowledge Discovery and Data Mining}} \emph{(\bibinfo{series}{KDD '22})}.
  \bibinfo{publisher}{Association for Computing Machinery},
  \bibinfo{address}{New York, NY, USA}, \bibinfo{pages}{4472--4481}.
\newblock
\showeprint[arxiv]{2206.06003}
\href{https://doi.org/10.1145/3534678.3539092}{doi:\nolinkurl{10.1145/3534678.3539092}}


\bibitem[Zhang et~al\mbox{.}(2023)]%
        {zhang2023leveraging}
\bibfield{author}{\bibinfo{person}{Yang Zhang}, \bibinfo{person}{Yimeng Bai},
  \bibinfo{person}{Jianxin Chang}, \bibinfo{person}{Xiaoxue Zang},
  \bibinfo{person}{Song Lu}, \bibinfo{person}{Jing Lu}, \bibinfo{person}{Fuli
  Feng}, \bibinfo{person}{Yanan Niu}, {and} \bibinfo{person}{Yang Song}.}
  \bibinfo{year}{2023}\natexlab{}.
\newblock \showarticletitle{Leveraging Watch-Time Feedback for Short-Video
  Recommendations: A Causal Labeling Framework}. In
  \bibinfo{booktitle}{\emph{Proceedings of the 32nd ACM International
  Conference on Information and Knowledge Management}}
  \emph{(\bibinfo{series}{CIKM '23})}. \bibinfo{publisher}{Association for
  Computing Machinery}, \bibinfo{address}{New York, NY, USA},
  \bibinfo{pages}{4952--4959}.
\newblock
\showeprint[arxiv]{2306.17426}
\href{https://doi.org/10.1145/3583780.3615483}{doi:\nolinkurl{10.1145/3583780.3615483}}


\bibitem[Zhao et~al\mbox{.}(2024)]%
        {zhao2024counteracting}
\bibfield{author}{\bibinfo{person}{Haiyuan Zhao}, \bibinfo{person}{Guohao Cai},
  \bibinfo{person}{Jieming Zhu}, \bibinfo{person}{Zhenhua Dong},
  \bibinfo{person}{Jun Xu}, {and} \bibinfo{person}{Ji-Rong Wen}.}
  \bibinfo{year}{2024}\natexlab{}.
\newblock \showarticletitle{Counteracting Duration Bias in Video Recommendation
  via Counterfactual Watch Time}. In \bibinfo{booktitle}{\emph{Proceedings of
  the 30th ACM SIGKDD Conference on Knowledge Discovery and Data Mining}}
  \emph{(\bibinfo{series}{KDD '24})}. \bibinfo{publisher}{Association for
  Computing Machinery}, \bibinfo{address}{New York, NY, USA},
  \bibinfo{pages}{4455--4466}.
\newblock
\showeprint[arxiv]{2406.07932}
\href{https://doi.org/10.1145/3637528.3671817}{doi:\nolinkurl{10.1145/3637528.3671817}}


\bibitem[Zhao et~al\mbox{.}(2023)]%
        {zhao2023uncovering}
\bibfield{author}{\bibinfo{person}{Haiyuan Zhao}, \bibinfo{person}{Lei Zhang},
  \bibinfo{person}{Jun Xu}, \bibinfo{person}{Guohao Cai},
  \bibinfo{person}{Zhenhua Dong}, {and} \bibinfo{person}{Ji-Rong Wen}.}
  \bibinfo{year}{2023}\natexlab{}.
\newblock \showarticletitle{Uncovering User Interest from Biased and Noised
  Watch Time in Video Recommendation}. In \bibinfo{booktitle}{\emph{Proceedings
  of the 17th ACM Conference on Recommender Systems}}
  \emph{(\bibinfo{series}{RecSys '23})}. \bibinfo{publisher}{Association for
  Computing Machinery}, \bibinfo{address}{New York, NY, USA},
  \bibinfo{pages}{528--539}.
\newblock
\showeprint[arxiv]{2308.08120}
\href{https://doi.org/10.1145/3604915.3608797}{doi:\nolinkurl{10.1145/3604915.3608797}}


\bibitem[Zhao et~al\mbox{.}(2019)]%
        {zhao2019recommending}
\bibfield{author}{\bibinfo{person}{Zhe Zhao}, \bibinfo{person}{Lichan Hong},
  \bibinfo{person}{Li Wei}, \bibinfo{person}{Jilin Chen},
  \bibinfo{person}{Aniruddh Nath}, \bibinfo{person}{Shawn Andrews},
  \bibinfo{person}{Aditee Kumthekar}, \bibinfo{person}{Maheswaran
  Sathiamoorthy}, \bibinfo{person}{Xinyang Yi}, {and} \bibinfo{person}{Ed~H.
  Chi}.} \bibinfo{year}{2019}\natexlab{}.
\newblock \showarticletitle{Recommending What Video to Watch Next: A Multitask
  Ranking System}. In \bibinfo{booktitle}{\emph{Proceedings of the 13th ACM
  Conference on Recommender Systems}} \emph{(\bibinfo{series}{RecSys '19})}.
  \bibinfo{publisher}{Association for Computing Machinery},
  \bibinfo{address}{New York, NY, USA}, \bibinfo{pages}{43--51}.
\newblock
\href{https://doi.org/10.1145/3298689.3346997}{doi:\nolinkurl{10.1145/3298689.3346997}}


\bibitem[Zheng et~al\mbox{.}(2022)]%
        {zheng2022dvr}
\bibfield{author}{\bibinfo{person}{Yu Zheng}, \bibinfo{person}{Chen Gao},
  \bibinfo{person}{Jingtao Ding}, \bibinfo{person}{Lingling Yi},
  \bibinfo{person}{Depeng Jin}, \bibinfo{person}{Yong Li}, {and}
  \bibinfo{person}{Meng Wang}.} \bibinfo{year}{2022}\natexlab{}.
\newblock \showarticletitle{{DVR}: Micro-Video Recommendation Optimizing
  Watch-Time-Gain under Duration Bias}. In
  \bibinfo{booktitle}{\emph{Proceedings of the 30th ACM International
  Conference on Multimedia}} \emph{(\bibinfo{series}{MM '22})}.
  \bibinfo{publisher}{Association for Computing Machinery},
  \bibinfo{address}{New York, NY, USA}, \bibinfo{pages}{334--345}.
\newblock
\href{https://doi.org/10.1145/3503161.3548428}{doi:\nolinkurl{10.1145/3503161.3548428}}


\end{thebibliography}

\end{document}